\documentclass[6pt]{article}
\usepackage[dvips]{graphics}
\usepackage{amsmath}
\usepackage{times}
\usepackage{xspace}
\usepackage{array}
\usepackage{epsfig}
\usepackage{latexsym}
\usepackage{amssymb}
\newcommand{\numub}{\ensuremath{\overline{\nu}_\mu}\xspace}
\newcommand{\nueb}{\ensuremath{\overline{\nu}_e}\xspace}
\newcommand{\numu}{\ensuremath{\nu_\mu}\xspace}
\newcommand{\nue}{\ensuremath{\nu_e}\xspace}

\newcommand{\nutau}{\ensuremath{\nu_\tau}\xspace}

\newcommand{\mum}{\ensuremath{\mu^{-}}\xspace}
\newcommand{\mup}{\ensuremath{\mu^{+}}\xspace}

\begin{document}
\vskip 0.5cm
\par \LARGE  Physics with charm particles produced  in neutrino 
interactions. A historical recollection.
\vskip 1cm
\Large

\par Ubaldo Dore 
\Large
\par

Dipartimento di Fisica, Universit\`a di Roma ``La Sapienza"\
\par
 I.N.F.N., Sezione di Roma, P. A. Moro 2, I-00185 Roma, Italy
\tableofcontents

\section{Introduction}
\label{intro}
\par This  paper presents
 a historical recollection of the neutrino experiments
 aimed at the study of charm physics in the last 40 years.
\par In the seventies, many experimental facts indicated the
 existence of new particles made by a new quark called "c" for charm. 
After the discovery of the J/Psi in 1974  \cite{augu,aube} clear-cut
 evidence for the c quark was expected
 from the measurement of the lifetime of particles with naked c. 
Lifetimes of the expected magnitude  were indeed first measured in 1976 
by a neutrino experiment using the high spatial resolution 
of emulsion detectors. Since then, neutrino experiments 
have continued to collect valuable information in different
 domains of charm physics. Because of the tiny neutrino cross-sections
huge detectors are needed and this  make direct observation difficult.
 Most of the experiments have identified charm production in their 
heavy detectors by tagging the production of leptons from charm decay. 
However, important results were also obtained by the direct 
observation of the decay of charmed particles in emulsion targets. 
\par This paper provides the basic information about all neutrino
 experiments relevant to the study of charm physics.
 We then summarize the present knowledge 
derived from these experiments on various issues,
 namely important between them the mass of the c quark, 
the s quark content of the nucleon, and the V$_{cd}$ and V$_{cs}$
  elements of the CKM matrix.
\par The paper is organized 
in the following way. First, in section \ref{disco}, 
we recall the discovery of charm, which marked the beginning of neutrino
 experiments in the field. Then, in section \ref{cross},
 we summarize the basic theoretical formalism nowadays in use
 to describe the process of charm production in high energy
 neutrino-nucleon scattering. 
A complete list of neutrino experiments on charm, together
 with their main characteristics is given in section \ref{experi}.
 The following sections contain a description of the experimental results.
 Results on opposite-sign dilepton events are summarized in section
 \ref{dimuon} where  the analysis of these events is described.
\par  Information on double charm production and on  multi-lepton events other
 than opposite-sign dilepton is discussed in section 
\ref{multi}.  Finally, section \ref{emul}, deals with 
 the few experiments which made use of nuclear emulsions 
to directly detect the short lived charmed particles.
Brief concluding remarks can be found in section \ref{conclu}.
\vskip 3cm

\section{The discovery of the charm quark}\label{disco}
The existence  of a fourth quark  was predicted by Glashow, 
Iliopoulos and Maiani \cite{maiani} in 1970. 
This quark was named charm by Bjorken  and Glashow \cite{Biork}
 and its  existence was experimentally proven by the discovery
 in 1974 of the J/$\psi$ \cite{augu,aube},the charm anti-charm meson 
state.
\par Naked charms, i.e. hadron states containing a charm quark 
 were first discovered in 1976~in~$e^{+} e^{-}$ annihilations at SLAC,
 by the observation of the D$^{+}$ \cite{peruzzi} and D$^{0}$
 \cite{piccolo,gold} mesons. 
The  characteristics  of the detected narrow states made
 the authors of \cite{peruzzi} write: 
 \par {\sl... strongly suggest  they are the predicted
 isodoublet (D$^{0}$ and D$^{+}$) of charm mesons.}
\par However, the definite conclusion that these particles
 were charmed hadrons required a determination
 of their lifetimes. In fact, the charm quantum number being
 conserved in strong and electromagnetic interactions,
 charmed hadrons were expected to decay weakly 
with lifetimes of the order of  $10^{-13}$ s. 
The  measurement of the lifetimes then required to use
 detectors with high spatial resolution, allowing to observe short
 decay paths.
\par In 1975 Marcello Conversi
 \cite{conversiprop} proposed to combine  emulsion, 
bubble chamber and counter techniques to study new
 shortlived particles produced by neutrinos. An experiment using spark
chambers and emulsions   was performed in 1976 at 
Fermilab \cite{Burhop,Read} and produced evidence for short-lived
particles produced in neutrino interactions. 
An experiment  conducted by Conversi, was then performed 
at CERN. Results were published in 1979 
 and succeeded to provide the first measurements of lifetimes 
for neutral and charged D particles \cite{conversi1,Allas4}. Charmed 
baryons were also observed  
  \cite{conversi2}.
These fundamental experiments opened the way to the study
 of charm physics with neutrino beams, subject of our paper. 
\par Before concluding this section, we should add a comment
 on the first observation of naked charm. We like to recall
 that 5 years before the discovery at SLAC, events later ascribed
 to associated production of charmed hadrons were observed 
in Japan in 1971 in studies of cosmic ray events
 with emulsion technique \cite{kniu}. A historical review on charm
physics with emulsions in Japan is given in 
 \cite{niu}.

\section{The mechanisms of charm production in neutrino nucleon scattering}
\label{cross}

\par We shall now outline the theoretical framework which is currently used to describe 
the process of charm production in neutrino nucleon scattering. We shall limit 
our discussion to the basic formulae which illustrate which are the quantities 
involved in the scattering process, and that can hence be studied by the experiments.
For a more detailed treatement of the theory and of the experimental techniques of charm 
physics with neutrinos we refer the reader to the review paper \cite{delelli}.
\par Charm particles are  produced mainly by neutrinos through the charged 
current (CC) reaction
$$ \numu+ N \rightarrow \mum +C+X $$.

\par where N is the target nucleon, C the charmed particle, X the remaining hadrons here 
and in the rest of this section we use as an example \numu scattering;
similar
relations hold for the \nue scattering.
 
\par Charm production on a fixed target requires neutrino energies above 
the threshold for production of the lightest charmed meson (m(D) = 1.8 GeV). Actually, 
the majority of neutrino experiments studying charm, use neutrino beams with 
energies from ten to a few hundreds GeV. In this energy domain the deep 
inelastic neutrino nucleon scattering is well described by the interaction 
of the neutrino with the quarks inside the nucleon. In this picture, the charm quark 
can be produced by a CC interaction on a constituent d quark, or on a d or s quark 
of the sea. To the lowest order, the scattering formula is written as \cite{nutev} 
$$ d^{2}\sigma_{c}(\numu \rightarrow \mum c X)/ dxdy =$$
$$2G^{2}M 
E_{\numu}/\pi[V^{2}_{cd}[((u(\xi,Q^{2})+d(\xi,Q^{2}))/2+V^{2}_{cs}s({\xi, Q^{2}})]$$
\par where $E_{\numu}$ is the neutrino energy, M the nucleon mass, $Q^{2}$ 
the 
four-momentum transfer squared, y the inelasticity, and x the Bjorken scaling 
variable (following the conventions given in \cite{steinbe}). The variable $\xi$ 
gives 
the actual fraction of nucleon momentum carried by the struck quark, taking into 
account kinematical effects, and is written as 
$$\xi=x(1+m_{c}^{2}/Q^{2})(1-x^{2}M^{2}/Q^{2})$$
where $m_{c}$ is the mass of the c quark. $V_{cd}^{2}$ and $V_{cs}^{2}$ 
are 
elements of the 
Cabibbo-Kobayashi-Maskawa  (CKM) quark mixing matrix. They take into account 
the two channels of production of the c quark, namely the 
transitions $d \rightarrow c$ and $s \rightarrow c$. The 
classification of quark in doublets, with the c and s quarks belonging to 
the same doublet makes $s \rightarrow c$ the natural transition for charm 
production. The transition $d \rightarrow c$ is also allowed, because of the 
mixing described by the Cabibbo angle. In neutrino 
scattering on nucleon both contributions represent only a few 
percent of the total cross-section. The transition $s \rightarrow c$ is 
disfavored because it involves the s quark, which is only present in 
the sea. The $d \rightarrow c$ channel is in turn suppressed by the 
smallness of the Cabibbo angle. 
\par A similar formalism can be used to describe $\numub$ nucleon 
scattering. An important difference is that with antineutrino the CC process 
leads to the production of the charm antiquark $\overline{c}$. The production of charm is then 
obtained from the scattering on $\overline{d}$ or $\overline{s}$, present only in the sea. 
\par Many corrections and details have to be studied, when applying to 
real data the formalism described above. For instance, the passage from quarks 
to the observed particles is usually parameterized in terms of the so 
called
fragmentation function D(z), where z is the fraction of the 
charm quark energy 
carried by the charmed hadron. The basic formulae given 
above clearly show how measurements of total and differential cross-sections 
for charm production with neutrino and antineutrino beams can give very important 
information on the structure functions and the structure of the nucleon, on 
the 
elements of the CKM quark mixing matrix, and on the mass of the c quark. Note 
that 
besides single charm production in CC interactions, other rarer processes have 
also been studied, like $c\overline{c}$ production in neutral current scattering
 (single c 
production can not happen by Z  exchange, since the Z  can not 
produce
a change of quark flavor).

\vskip 3 cm

\section{Experiments on charm physics with neutrino beams}
\label{experi}

\par Charm physics has been studied with high energy neutrino beams by many different 
experiments at Fermilab (Fermi National Accelerator Laboratory) and at CERN (European
Organization for Nuclear Research). Most of the experiments made use of general purpose 
neutrino detectors. Because of the smallness of the neutrino cross-sections, these 
detectors are very massive and serve at the same time as target. They usually have limited 
granularity and do not allow to identify single hadrons. Measurements of 
hadrons 
are in this case restricted to the energy and direction of the hadronic shower. Charged muons 
instead, are penetrating particles and can therefore be identified, and their momentum 
be measured in a magnetic field. The usual event classification 
in CC or NC 
(Charged or Neutral Current) interactions is then  based, for \numu interactions for example, on the presence 
or absence of a \mum in the final state. With this kind of detector, charm physics could be 
studied exploiting the semileptonic decay of the c quark, $c \rightarrow s(d) + \mup + \numu $. A 
 typical example is the reaction 
$$ \numu + N \rightarrow \mum + C + X $$
\par where C is a charmed hadron and X the system of the remaining hadrons. C can decay with 
a substantial branching ratio producing a \mup. The final state is then characterized by the 
simultaneous presence of a \mum and a \mup. Events of this kind  are called
 $$\sl{"opposite~sign~dimuons"}$$
 and
constitute the largest sample of charm events studied with neutrino beams. A background is present,
mainly due to pions of the hadronic shower, which occasionally decay or simulate the behaviour of
muons. Nevertheless, statistical analysis of the data have yielded very accurate  measurements of
differential cross-sections for charm production, and allowed the studies mentioned in the previous
section.  More in general, the identification of muons, and in some case of electrons, has been
the main tool of investigation of charm physics with neutrino beams. Note however that other important
results were obtained by a few experiments using emulsion targets. In this case the high spatial
resolution allowed to directly identify charmed hadrons through the observation of decay paths of
few millimeters. Then, properties of the charmed hadrons, like lifetimes and branching ratios
could be directly investigated.

\par In this section, after briefly recalling how neutrino beams are produced, we list all 
experiments that provided significant results on charm physics, giving a short 
description 
for each of them.  Physics results obtained by the experiments will then be illustrated and 
summarized in the following sections 5 to 7.

\subsection{ Neutrino beams }

\par Neutrino beams for charm studies were operated at CERN and at Fermilab using their respective 
proton accelerators. Different setups were  adopted to vary the energy 
spectrum and flavour content
of the beam, but the basic principle of operation is always the same. To illustrate it, we take 
the example of the Wide Band Beam (WBB) generated with the 400 GeV Super Proton Synchrotron (SPS) at CERN. 

\begin{figure}[hbtp!]
\begin{center}
\epsfig{figure=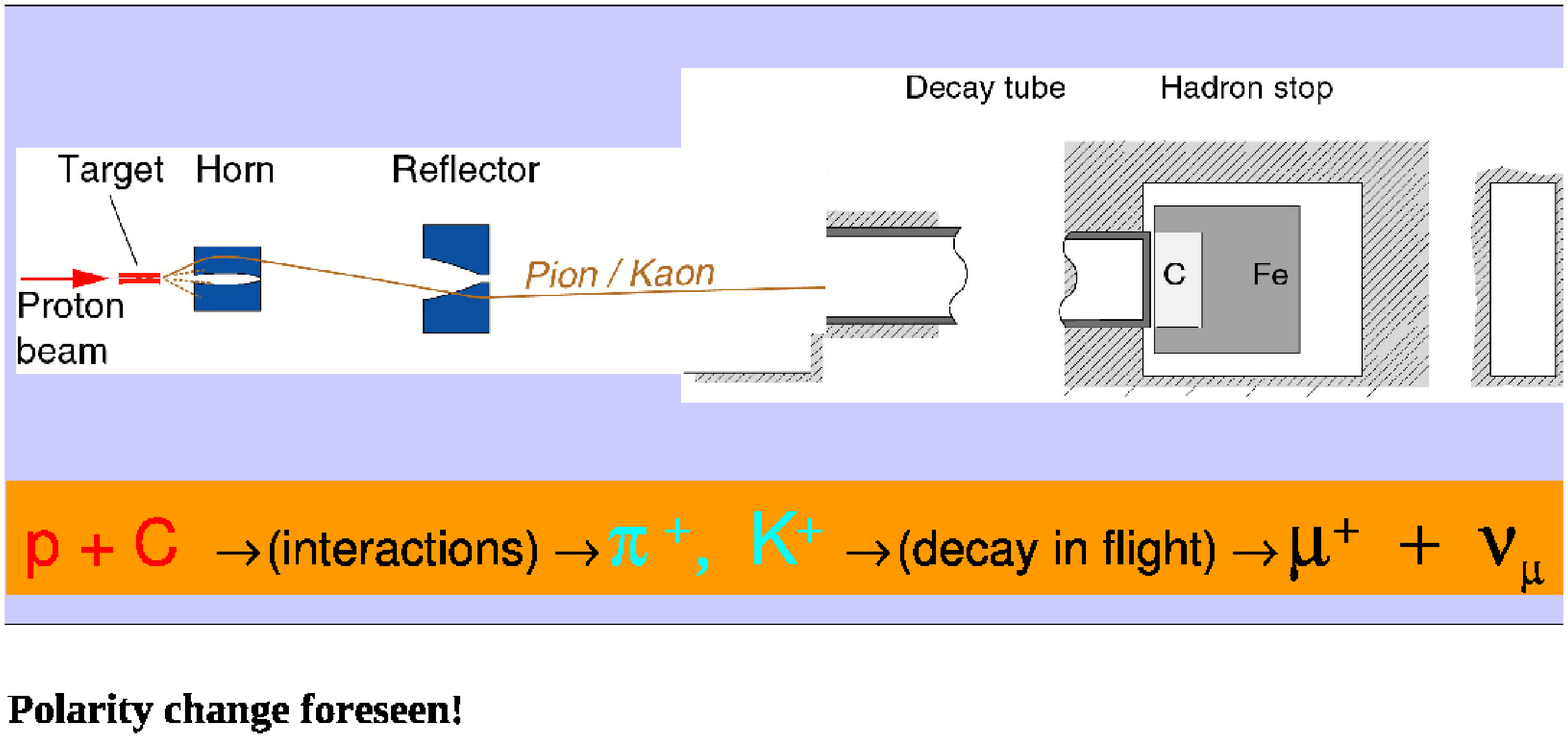,width=14.cm}

\caption {Setup of the CERN Wide Band neutrino Beam (WBB) }
\label{beam}
\end{center}
\end{figure}

A scheme of the beam layout is shown in figure \ref {beam} (sizes and distances are not in 
scale). Protons extracted
from the accelerator impinge on a cylindrical target where they interact producing pions and kaons. Two magnets
with specially designed field configuration, the so-called Horn and 
Reflector, allow to focus and charge-select 
 the mesons. The mesons go through the decay tunnel, an evacuated region 258  meters long,
where a substantial fraction of them  decays producing a collimated beam of neutrinos (neutrinos for positive
mesons, antineutrinos for negative mesons). After the decay tunnel, the remaining hadrons and the muons
that are produced along with the neutrinos are absorbed in a shield of adequate thickness. For given intensity
and energy of the primary proton beam, the characteristics of the neutrino beam (intensity, energy spectrum,
flavor content) depend  from the design of the focusing magnetic field and from the length of the decay tunnel.
The neutrino energy spectrum of the CERN WBB peaks typically around 30 GeV, and the intensity drops
by a factor 100 at about 150 GeV. The total contamination of wrong flavours
($\numub, \nue, \nueb)$ is
only a few percent in the \numu beam, and goes up to 10$\%$  in the \numub beam. At 
CERN it was also
possible to run the so-called Narrow Band Beam (NBB), which operated a selection in momentum of pion and
kaons (typically at 200 GeV). The NBB had much lower intensity but provided a two-components (pions and kaons) flat
neutrino energy spectrum and better control of the contamination of neutrinos of the wrong flavor.
\par Similar techniques were adopted at Fermilab where the proton beam of the Tevatron accelerator ran  with energies 
up to 1000 GeV. A list of the Fermilab neutrino beams can be found in ref \cite{kopp}.

\subsection{ List of experiments}

\par In the following we list, first for Fermilab, then for CERN, all experiments that provided significant
results on charm physics. For each experiment we give schematically the characteristics of the detector,
the year of first data taking, and a  list of  the main studies performed, other than those on charm. Reference
is made to some of the most representative publications.

\subsubsection{ Fermilab experiments}

\begin{itemize}

\item \underline {E247 (1976)} 
\par The E247 experiment consisted of an emulsion target followed by a set 
of spark chambers. It took data in the wide band neutrino beam produced by 
600 GeV protons. It produced the first evidence for the production of 
charmed hadrons in neutrino interactions  \cite{Burhop}.

\item \underline {HPWF ( 1974)}
\par The detector of the HPWF collaboration \cite{benve6} consisted of an iron 
target, a liquid scintillator calorimeter \cite{benve2} and an iron-plates 
muon identifier \cite{benve1}. The experiment ran in the 
quadrupole-triplet and in the
bare-target-sign-selected  beams at Fermilab  in the seventies. In 1975 it made the first
observation of dimuon events  \cite{aubert}. It must be noted that in 1974 the collaboration
made one of the first confirmation of the existence of 
neutral currents \cite{benve8}.

\item 
\underline{CCFR, E616, E701, E744, E760 ( 1976)}
\par In 1976 the E616 experiment initiated by the collaboration of  
the California Institute of Technology, the Chicago University, Fermilab, 
the University of Rochester and the Rockfeller University,
was the first to use high energy neutrinos (energies up to 300 GeV) in the  
LabE at  Fermilab. The calorimeter of the Collaboration was equipped with spark chambers
that limited the data acquisition to one event per extraction. Measurements 
of neutral current electroweak parameters were reported by experiments E616 
and then E701 \cite{Reutens}.
\par The collaboration then became the CCFR collaboration (California, Columbia,
Fermilab and Rochester).
The CCFR detector consisted of on heavy target calorimeter, a muon 
spectrometer and  drift chambers \cite {sakumoto} that 
replaced the E616 spark chambers.
The E744 and E770 experiments took data in 1987-1988 in the neutrino beam from the 
Tevatron (neutrino energies up to 600 Gev). Main results refer to: 
\begin{itemize}
\item structure functions \cite{selig}
\item strange quark  content of the nucleon \cite {bazarko}
\item neutral currents \cite {arroyo}
\item dimuons \cite{foud}
\end{itemize}
\par Some of these results have been obtained together with the NuTeV
collaboration

\item \underline{15-foot bubble chamber (1974)}
\par The so-called 15-foot bubble chamber was at the time of construction 
the largest liquid hydrogen bubble chamber. It started its  operation in 1974.
In 1987-1988 the E632 Collaboration used the 15-foot chamber,
filled with a  neon-hydrogen mixture and equipped with an 
external muon identifier. E632 took data with the quadrupole-triplet 
neutrino beam at the Tevatron and was the
only bubble chamber experiment to study 
neutrino interactions at Tevatron energies. Studies were made on:
\begin{itemize}  
 \item dimuons \cite{ballag}
 \item neutral currents \cite{Ader}
 \item strange  particles production \cite{Depro}  
 \item holography \cite{hari}

\end{itemize}

\item \underline {E531 (1981)}
\par The E531 experiment was an hybrid emulsion experiment made of  a 23 liter   
emulsion target, a large aperture analysis magnet, drift chambers, time of flight 
hodoscopes for charged particles identification, a lead glass array and  
an electromagnetic  calorimeter followed by a muon identifier.  
The experiment ran in the eighties at Fermilab in the wide band neutrino beam 
produced by 350 GeV protons and peaked at an energy of 25-30 GeV. The 
experiment was 
designed to study the production and decay of short lived particles produced in neutrino 
interactions \cite{e531} and  also searched for 
$\numu \rightarrow \nutau$ \cite {ushida} oscillations.

\item \underline{ NuTeV ( 1996)}
\par
NuTeV (E815) was an upgraded version of the CCFR detector. It consisted of 
an 
iron scintillator sampling calorimeter interspersed with  drift chambers and
followed my a muon toroidal magnet \cite{nutevcal}. The experiment took data in
1996-2001. NuTeV was the last experiment to run in the Fermilab  LabE.
Results have been obtained on:
\begin {itemize} 
\item dimuons \cite{nutev}
\item neutrino cross sections \cite{boyd} \cite{bodek}
\item oscillations \cite{avvaku}
\item sin$^{2} \theta_W $ \cite{zeller}
\end{itemize} 
\end{itemize}

\subsubsection{CERN experiments}

\begin{itemize}

\item  \underline{ Gargamelle (1970)}
\par The Gargamelle bubble chamber was built in the Saclay laboratory in 
France. Filled with propane or freon it was operated from 1970 to 1978  
in the CERN neutrino beam. In 1972 data collected with Gargamelle
led to the discovery of neutrino neutral current interactions 
\cite{neutral}, providing a fundamental step in the understanding  
of particle physics. Many important results refer to: 
\begin{itemize}
\item dileptons \cite{garga2}
\item structure functions \cite{allas1}
\item cross sections \cite{allas3}
\item neutral current \cite{allas2}
\item beam dump experiment \cite{Alibran}
\item production of strange particles \cite {Erriquez} 
\item   neutrino oscillations \cite{arme2}.
\end{itemize}

\item 
\underline {BEBC (1973)}
\par The Big European Bubble Chamber (BEBC) was installed
at CERN in the early 70' and  was dismantled in 1985. The vessel was filled with 35 
cubic meters of hydrogen or deuterium or neon-hydrogen mixture. The 
addition of a external muon identifier, an external particle identifier, an 
internal picket fence and a track sensitive target later converted BEBC in a hybrid 
detector. BEBC took data from 1973 in a PS beam then from 1977 to 1984 in a SPS beam
and concluded his life with a PS oscillation experiments \cite{Angeli}.
\par In 1979 for the first time decays of charmed particles were observed in a bubble chamber
\cite{conversi1,conversi2,Allas4}.
A complete list of BEBC results can be found in \cite{hari2}.
Out of the many studies dedicated to neutrino physics, we recall:
\begin{itemize}
 \item measurements of neutrino cross sections \cite{ bosetti} 
 \item oscillation experiments \cite{errique}
 \item dimuons \cite{arme1}
 \item beam-dump \cite{grass1}
\end{itemize}

\item \underline {Conversi (1976)}
\par The hybrid experiment of Conversi at CERN consisted of an emulsion target, followed
by a bubble chamber and various electronic detector. The experiment ran in the 
CERN wide band PS  neutrino beam and observed in emulsion the decay of 
charmed 
hadrons,
providing the first measurements of lifetimes 
for neutral and charged D particles \cite{conversi1,Allas4}.

\item
\underline {CDHS  (1977)}
\par The CDHS collaboration  (CERN,Dortmund,Heidelberg, Saclay), 
 initially led by J. Steinberger, took data from 1977 in the 
neutrino beam of the CERN SPS. In 1983 the CDHS detector was split in two parts
and used for a neutrino oscillation experiment in the PS neutrino beam \cite{Dydak}.
\par The detector \cite{cdhsdet}   consisted of circular magnetized iron 
plates, interspersed with drift chambers, for a total mass of 1250 tonnes. 
\par In 35 publications the experiment made contributions in the following
fields: 
\begin{itemize}
\item neutrino and anti-neutrino cross-sections \cite{degrot} 
\item structure funtions \cite{degrott}
\item beam-dump experiment \cite{Hansl}
\item CKM mixing matrix \cite{abra} 
\item electroweak parameters \cite{blondel}
\item multimuon events \cite{abra,degro,trimucdhs}
\item neutrino oscillations \cite{Dydak}
\end{itemize}   

\item
\underline{CHARM  (1978)}
\par The CHARM (CERN, Hamburg, Amsterdam, Rome, Moscow) Collaboration 
started to operate his detector in 1978, taking data with the CERN 
SPS neutrino beam. The detector was a low density, low Z calorimeter optimized for 
the detection of neutral currents. It consisted of 78 marble plates 
interspersed with planes of proportional tubes and scintillators. 
The marble plates were surrounded by magnetized iron 
frames. The calorimeter was followed by a muon spectrometer 
with magnetized iron toroids \cite{Diddens}.
\par Main results refer to:
\begin {itemize}
 \item structure functions \cite {berg2}
 \item neutrino electron scattering \cite{doren}
 \item beam-dump experiment  \cite{allab1}
 \item polarization of positive muons \cite{jonk1}
 \item  neutrino oscillations \cite{bergsma}
 \item the electroweak mixing angle \cite{allab2}
\end{itemize}

\item 
\underline{CHARMII (1982)}
\par The CHARMII detector was designed to study neutrino electron scattering.
It ran in the CERN SPS neutrino beam. It was a massive fine grained 
calorimeter (692 t) of low density (target material glass) 
followed by a magnetic spectrometer \cite{charm2det}.
\par Results were obtained on: 
\begin{itemize}
 \item neutrino and antineutrino electron scattering and determination of the
 electroweak mixing angle \cite{vilain4}
 \item inverse muon decay \cite{vilain1}
 \item search for $\numu \rightarrow \nutau$ oscillations \cite{gruwe}
 \item  search for muon to electron oscillations \cite{vilain3} 
 \item QCD analysis of dimuons events \cite{vilain2}
\end{itemize}

\item
\underline{NOMAD (1994)}
\par 
The NOMAD experiment \cite{nomaddet} was designed to detect $\numu \rightarrow \nutau$
oscillations. It ran in the CERN SPS neutrino beam. The active target consisted of a set 
of drift chambers (target mass 2.7 ton) located in a magnetic field  (0.4 T).  The active target  
was followed by a transition radiation detector to identify electrons, by an electromagnetic and 
a hadronic calorimeter, and  by muon chambers.
Results were obtained on: 
\begin{itemize}
 \item neutrino cross-sections \cite{Qwu}
 \item neutrino oscillations \cite{aste2}
 \item dileptons \cite{aste}
 \item production of strange particles \cite{naumov} 
 \item meson resonances production \cite{astie4}
 \item D* production \cite{astie3}
\end{itemize}

\item
\underline{ CHORUS (1994)}
\par The CHORUS experiment took data from 1994 to 1997 in the SPS CERN neutrino 
beam. The detector \cite{Eskut2} was specially designed to detect 
$\nu_\mu \rightarrow \nu_\tau$  oscillations. It was a hybrid detector with a 
large (770Kg) emulsion target
followed by planes of high resolution fiber trackers used to predict the position
in the emulsion of the neutrino 
interaction vertex by extrapolating backwards particle tracks. 
\par Limits on $\nu_\mu \rightarrow \nu_\tau$  oscillations are published in 
\cite{Eskut1}. The results on charm physics will be described in section
 \ref{emul}.

\end{itemize}
\vskip 3 cm

\section{ Charm physics with opposite-sign dimuons }
\label{dimuon}

\subsection{Introduction}

\par Opposite-sign dimuons have been the main tool of investigation of charm 
physics with neutrino beams. In that process, described in the previous 
sections, the leading muon (the muon of highest energy) is interpreted 
as originating from the neutrino vertex, while the opposite-sign muon 
(or electron in bubble chamber detectors) as due to  the leptonic  
decay of the charm inside the hadronic shower.

\par The history of dimuons starts already in 1974, the year of 
the J/Psi discovery, when the HPWF Collaboration at Fermilab found 
two events attributed to the reaction 
$\numu+N \rightarrow \mu+hadrons$ 
followed by a muonic decay inside the hadronic shower. 
In their paper the authors mention the following 
interpretation \cite{aubert}: 
{\sl One possibility is that the particle jet produced 
by very high energy neutrinos has an anomalously large 
probability to decay into a (positive) muon. Models 
generating this type of effect due to the presence of novel 
hadronic quantum numbers (charm) have been recently 
suggested by Glashow, Iliopoulos and Maiani \cite{maiani}}. 
\par The HPWF experiment was also the first to confirm in 
1975 charm production in a neutrino beam by publishing 
results obtained with a sample of about 100 dimuons \cite{benve4}. 
From then on, all neutrino experiments searched for dimuon events 
and a complete list of dimuon (or better dilepton) experiments 
is given in the following table.

\begin{table}[htbp]
\begin {center}
\begin{tabular}{|c|c|c|c|c|c|}\hline
 n &{\bf Experiment} & authors &reference & technique\\\hline
 1&HPWF& A. Benvenuti et al. 1988&\cite{benve5}& counter exp\\
2&BEBC &G. Gerbier et al. 1985& \cite{Gerbier}&bubble chamber\\
3&Gargamelle&N.Armenise et al. 1979, &\cite{garga1}&bubble chamber\\
4& Gargamelle& A. Haatuft et a.1983& \cite{garga2}& bubble chamber\\
5& LBL coll&H.C. Ballagh et al.1981&\cite{ballag}& bubble chamber\\
7&CB coll.& N.J. Baker et al.1985&\cite{baker}&bubble chamber\\
6& CDHS coll.&Abramowicz et al. 1982&\cite{abra}& counter exp\\
9& E616& K.Lang et al. 1987&\cite{lang} &counter exp\\
10& CCFR&A.Rabinowitz et al. 1993&\cite{ccfr}&counter exp \\
11&NUTEV&M. Goncharov et al. 2001&\cite{nutev}&counter exp\\
11& CHARMII & P. Vilain et al.1999 &\cite{vilain} & counter exp\\
12& CHORUS &A.Kays-Topaksu et al. 2008& \cite{lead}& counter exp\\
13&NOMAD& P. AStier et al. 2000 &\cite{aste}& counter exp\\\hline
\end{tabular}
\caption {\large{Dilepton  experiments}}
\label {expe}
\end {center}

\end  {table}
 
\par The CDHS Collaboration at CERN reached in the early eighties 
the largest statistics, with 9922 dimuons collected in neutrino 
beam, and 3123 in antineutrino beam. The picture was later completed 
with the high energy data collected in 2001 by the NuTeV 
experiment \cite{nutev} using the neutrino beam produced 
by the 1 TeV protons of the Fermilab Tevatron.
\par
\begin{figure}[hbtp!]
\begin{center}
\epsfig{figure=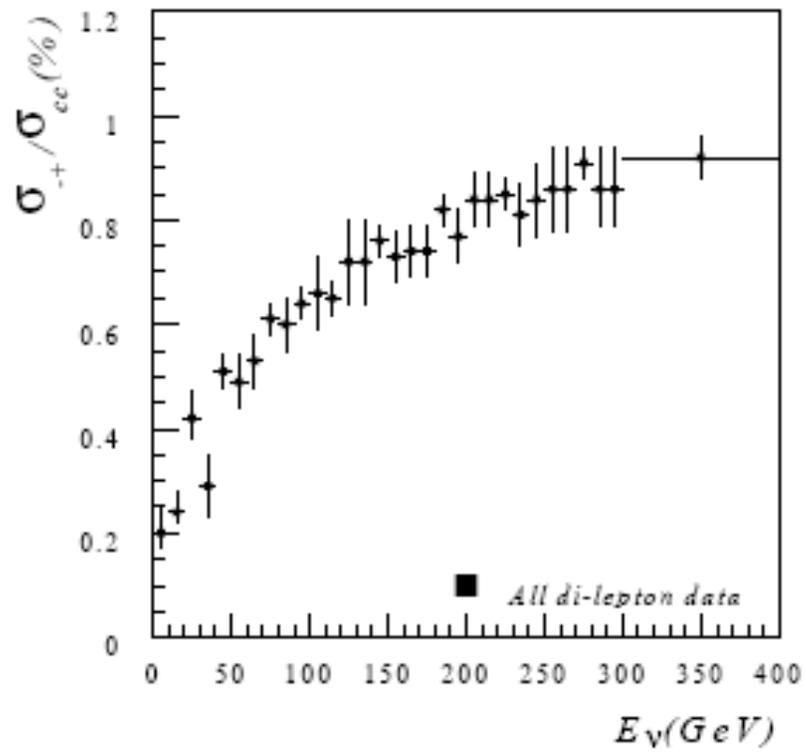,width=12.cm}\caption {Neutrino 
dimuon cross-section from \cite{migliozzi}}
\label{crosss}
\end{center}
\end{figure}

 The large statistics collected at various energies 
in \numu and \numub beams allowed to investigate in detail 
the production mechanism of charm outlined in section 3. 
Quality and statistics of the data collected by the various 
experiments between 1975 and 2000 are well 
illustrated by the plot shown in Fig. \ref{crosss}.
 
\par The figure comes from a review paper of 2002 \cite{migliozzi} 
where the measurements of the opposite-sign dimuon 
cross-section $\sigma_{+-}$ are combined in a single plot. 
The plot gives the ratio of $\sigma_{+-}$ to 
the CC cross-section for \numu N scattering as a 
function of $E_{\numu}$. At low energy it shows with high 
precision the kinematical suppression generated by the large 
mass of the c quark. At high energy it gives the plateau 
value of 0.9$\%$ and shows that the c quark contribution to 
the CC cross-section behaves similarly to all other hadronic 
processes. Note however  that the review did not include the 
high energy data of NuTeV \cite{nutev}.
\par  Dimuons data taken with neutrino and antineutrino 
beams were extensively analyzed in order to extract information 
on the process of charm production. The results obtained for 
each of the physics properties involved in the scattering 
process will now be presented and discussed.

\subsection{The analysis of dimuon events}
\par The basic formalism adopted by the experiments to analyze 
dimuons was discussed in section \ref{cross}, where the QCD 
lowest order cross-section for charm production is given. 
Then, when analyzing dimuons, one has to take into account 
the fragmentation of the quark into hadrons and the branching 
ratio $B_{\mu}$ for the inclusive muonic decay of the charmed 
hadron. The usual description has the form 
\par $$ d^{3}(\sigma_{+-}) / d\xi dy dz=
 d^{2}\sigma_{c}(\numu N \rightarrow cX)D(z)B_{\mu}/d\xi dy$$
\par Here D(z) is the fragmentation function for the charm quark,
with z=E$_{C}$/E(hadrons) defined as the fraction of
the total hadronic energy taken by the charmed hadron. 
D(z) and  $B_{\mu}$ are usually both averaged over all 
charmed hadrons produced.
\par The basic formalism is then used to fit with 
MonteCarlo techniques the experimental distributions generally 
obtained from the measurements of the momenta of the two muons, 
of the energy and direction of the hadron shower, and from the 
known direction of the incoming neutrino. Similar analysis techniques 
were used by all experiments and their various results are summarized 
hereafter.

\begin{itemize}
\item \underline{$m_{c}$, the mass of the charm quark}
\par Unlike leptons, quarks are confined inside hadrons and are not 
observable as free particles. Therefore, quark masses cannot 
be directly measured and must be defined within a theoretical 
framework. However, quantitative kinematical effects are produced 
by the mass of the charm quark, which in particular generates 
threshold effects on the dimuon cross-section. This effect 
is usually taken into account using the so-called slow rescaling 
formalism \cite{georgi}. Fig \ref{corre} shows the effect of $m_{c}$, the 
c quark 
mass, on the low energy behavior of cross sections, and the 
correction obtained by the CCFR collaboration using a leading 
order low-rescaling formalism \cite{ccfr}.

\begin{figure}[hbtp!]
\begin{center}
\epsfig{figure=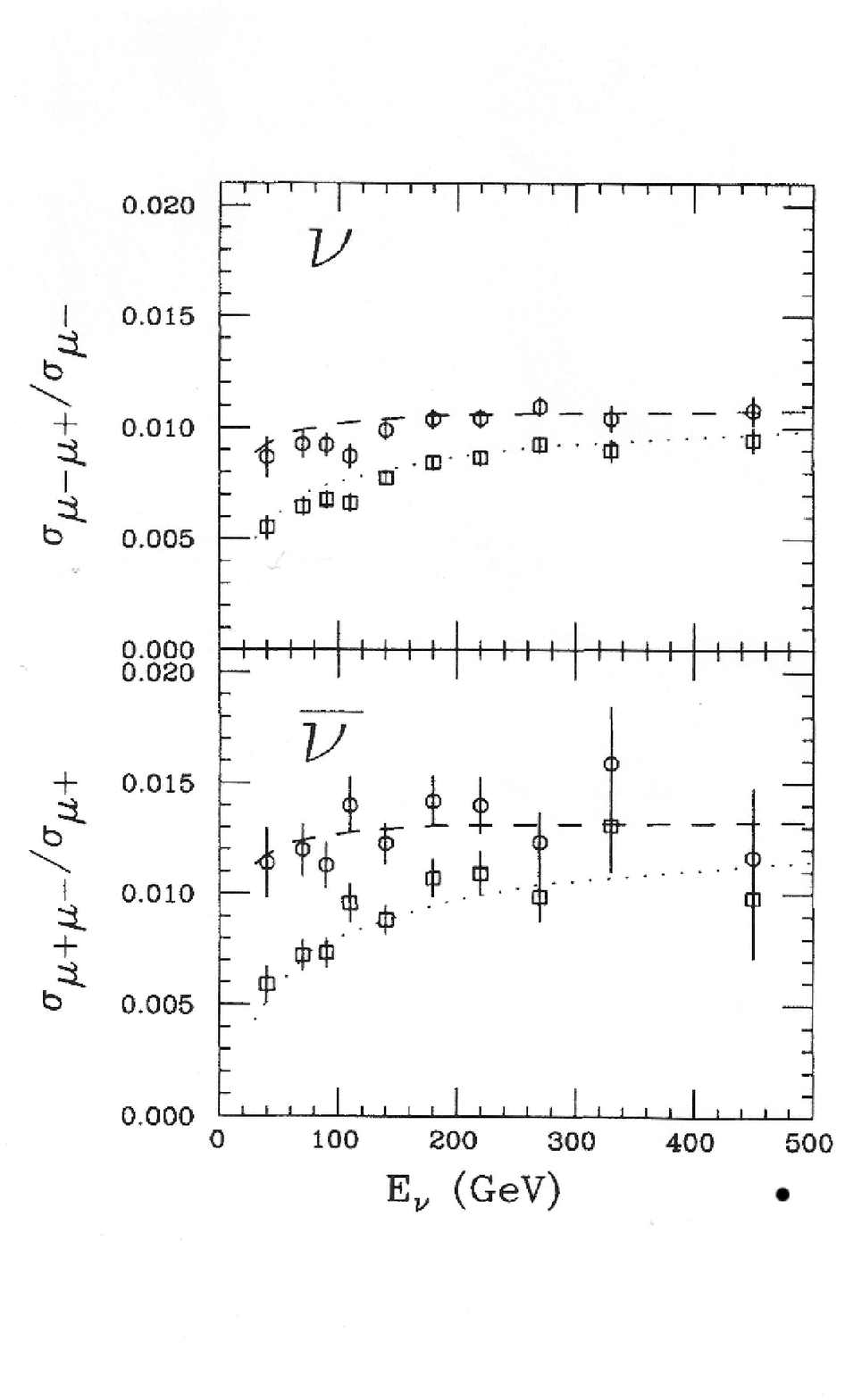,width=10.cm}
\end{center}
\caption {Neutrino and antineutrino cross-section 
ratios from ref \cite{ccfr}. Squares: data. Circles: data 
corrected with the low rescaling formalism.}
\label{corre}
\end{figure}

\par Values of $m_{c}$ obtained by different experiments 
using the slow-rescaling formalism are given in Table \ref{para}.

\begin{table}[htbp]
\begin{center}
\begin{tabular}{|c|c|c|c|c|c|}\hline
Experiment &N$_{2\mu}$($ \numu)$& N$_{2\mu}$($\numub)$& m$_{c}$(GEV)&
$\sl{k}$&B$_{\mu}$\\\hline
CDHS \cite{abra}& 9922 & 3123&1.26 $\pm$ 0.18& 0.52 $\pm$ 0.09 &0.071
$\pm$0.013\\
CCFR \cite{ccfr} & 5044& 1062&1.31$^{+0.20+0.12}_{-0.22-0.11}$& 0.373
$\pm$
0.09&0.084 $\pm$ 0.014\\
CHARMII \cite{vilain}& 3100 &700 &1.8 $\pm$0.4 &0.39 $\pm$ 0.09 &0.091
$\pm$0.010\\
NOMAD \cite{aste}& 2714 &655 &1.3 $\pm$ 0.4& 0.48 $\pm$ 0.17 &0.095 $\pm$
0.015\\
NUTEV \cite{nute2}&5102& 1458&1.24 $\pm$ 0.25 &0.42 $\pm$ 0.08&0.101 $\pm$
0.012\\
CHORUS\cite{lead}&8910 & 430& 1.26 $\pm$ 0.18&0.33 $\pm$0.07 &0.096 $\pm$ 0.008\\\hline
\end{tabular}
\caption{Parameters determined from dimuon events}
\label {para}
\end{center}
\end{table}

\par Attempts have been made to determine the value 
of $m_{c}$ to the  next to leading order. For example, 
the CCFR Collaboration with such analysis quotes \cite{bazarko} 
a value of 1.6 $\pm$ 0.16 GeV for $m_{c}$, while at the leading 
order the same Collaboration gets $m_{c}$ = 1.31  GeV (see table \ref{para}).
\par It is worth mentioning that the value of $m_{c}$ enters also 
in the determination of sin$^{2}\vartheta_w$ from the ratio NC/CC in 
neutrino-nucleon scattering. For example, the CHARM 
Collaboration \cite{charmtet} quotes his measurement in the form
\par   sin$^{2}\vartheta_w$ =0.234+0.012( $m_{c}$-1.5 GeV) $\pm$0.0051
$\pm$0.0024.

\item
\underline{ $\sl{k}$: the strange quark content of the nucleon}
\par Since the charm quark is produced in neutrino interactions 
also by the scattering off the s quark, which is only present 
in the sea, the study of dimuon differential cross-sections can 
be used to determine the strange quark content of the nucleon. 
Neutrino experiments usually give results for $\sl{k}$, where $\sl{k}$ 
is defined as the ratio of strange quark over 
sea $\overline{d}$ and $\overline{u}$ antiquarks,
\par $\sl{k}$=2s/($\overline{d}+\overline{u}$).
\par The values of $\sl{k}$ obtained by various experiments 
are given in table \ref{para}. They agree with each other 
and show a marked asymmetry of the strange quark sea 
compared to the $\overline{d}+\overline{u}$ average.
\par  The CCFR Collaboration \cite{ccfr} has also studied:
\par a) the ratio of the strange quark distribution s over 
that of the  valence quark u+d. They 
find $\eta_(s)= 2s/(u+d)=0.064 \pm 0.008 \pm 0.002 $
\par b) the x dependence of the sea 
quark defined as
$$ s(x)=(1-x)^{\beta} / x $$
\par The obtained value, $\beta =9.45 $, shows that the strange
sea is softer than the total sea ($\beta = 6.95$).
\par It must be noted that no difference between s and $\overline{s}$
distributions has been found \cite{bazarko}.

\item \underline{$B_{\mu}$: the muonic charm branching ratio}
\par $B_{\mu}$, the average inclusive muonic branching ratio 
of charmed hadrons, is usually treated in the full analysis 
of the dimuons differential cross-sections as an overall 
normalization factor. Values of $B_{\mu}$ obtained by the 
various dimuon experiments are given in table \ref{para}. 
There is also a direct measurement of the muonic 
branching ratios performed in emulsions by the CHORUS 
experiment. This will be discussed in section \ref{emul}).

\item \underline{ The fragmentation of the charm quark }
\par The hadronization of the charm quark is usually described 
by a phenomenological fragmentation function D(z)(z=E$_{C}$/E(hadrons)).
Many neutrino experiments obtained a satisfactory description 
of the data using for D(z) the Peterson parameterization\cite{peter} 
$$D(z)=[1/z(1-1/z-\epsilon_{p}
/(1-z)]^{-2}$$
\par where $\epsilon_{p}$ is a parameter to be determined 
by the fit to the dimuons distributions. Results of the 
fits performed by the various experiments are shown 
in table \ref{epsil} by quoting the value of the 
average z and, for those experiments adopting the Peterson 
parameterization,  the value of $\epsilon_{p}$.

\begin{table}[htbp]
\begin{center}

\begin{tabular}{|c|c|c|c|}\hline
Experiment&z average& $\epsilon_{p}$\\\hline
BEBC\cite{asra}& 0.59 $\pm$0.09&\\
E531 \cite{e531c}&0.61$\pm$0.02&0.076$\pm$0.014\\
CHORUS calorimeter \cite{lead}&0.61$\pm$.05&0.040$\pm$0.015\\
CHORUS emulsions(*)
\cite{frag}&0.63$\pm$0.103&0.108$\pm$0.017$\pm$0.013\\
NOMAD\cite{astie3}&0.67$\pm$0.03&0.075$\pm$0.046\\
CDHS\cite{abra}&0.68$\pm$0.08&\\
CCFR \cite{ccfr}&0.56$\pm$0.03&0.22$\pm$0.05\\
CHARMII \cite{vilain} &0.66$\pm$0.03&0.072$\pm$0.017\\\hline
\end{tabular}
\caption{Values of z average and  $\epsilon_{p}$ .
 (*) for Chorus emulsions see section \ref{emul}}
\label{epsil}

\end{center}
\end{table}

\par Other fragmentation properties of charm have 
been obtained by NOMAD, studying the  D$^{+*}$ production \cite{astie3}, 
by E531 \cite{e531} from the study of inclusive charm production, 
by BEBC from the study of D$^{+*}$ \cite{asra} 
and by CHORUS from its large sample of D$^{0}$ \cite{D02}.

\item \underline{Measurements of the CKM matrix 
elements $V_{cd}$ and  $V_{cs}$}
\par The cross-section for charm production 
in neutrino and antineutrino scattering on nucleons 
is directly related to the relative strength of the coupling 
of the c quark with the s and d quarks. Fits to the dimuons 
differential cross-sections have therefore been used 
to determine the $V_{cd}$ and  $V_{cs}$ elements of the 
CKM matrix. These matrix elements have been measured in many 
different reactions and a recent review of all measurements, 
including those from neutrino experiments, can be found 
in ref \cite{datagr1}. It turns out that dimuons give the 
best overall determination of $V_{cd}$, while the 
measurement of $V_{cs}$ in neutrino scattering suffers 
from the uncertainties on the sea quark content of the 
nucleon. The results on $V_{cd}$ and  $V_{cs}$ from the 
analysis  of dimuons are the following.

\par \underline {Measurements of $V_{cd}$}
\par The determination of $V_{cd}$ is derived from 
the difference of the dimuon over single-muon cross-section 
ratios for neutrinos and antineutrinos, a difference 
which is proportional to $B_{\mu} V_{cd}^{2}$. This 
method was first used in 1982 by the CDHS Collaboration, 
and then adopted in all neutrino experiments. The results 
of their analyses are given in table \ref{vcd}.

\begin{table}[htbp]
\begin{center}

\begin{tabular}{|c|c|c|c|c|}\hline

Experiment &  $ B_{\mu}$$V_{cd}^{2}$ & $ B_{\mu}$&$V_{cd}$ &ref\\\hline
CDHS & 4.11$\pm$0.07 .10$^{-3}$ &7.1$\pm$1.3 10$^{-2}$ &0.24$\pm$0.03&
 \cite{abra}\\\hline
CCFR
&5.34$^{+0.038+0.27+0.25}_{-.39-0.21-0.05}$.10$^{-3}$&9.9$\pm$1.2
.10$^{-2}$&0.232$^{+0.018}_{-0.020}$

&\cite{bazarko}\\\hline
 CHARMII&4.75$\pm$0.27 .10$^{-3}$ &  9.1 $\pm$1.0 10.$^{-2}$&0.227$\pm
0.006$ $\pm$
0.011&\cite{vilain}\\\hline
CHORUS&04.74 $\pm$ 0.27$^{-3}$ & 9.6 $\pm$0.8 10.$^{-2}$ &0.222
$\pm$0.016&
 \cite{lead}\\\hline
\end{tabular}
\caption{$V_{cd}$ values }
\label{vcd}
\end{center}
\end{table}
\par A combined analysis of the results from dimuons shown 
in table \ref{vcd} gives V$_{cd}$=0.230$\pm$0.011 \cite{datagr1}, 
which represents the best determination of V$_{cd}$, 
compared to those obtained in any other sector.  

\par \underline {Measurements of $V_{cs}$}
\par Measurements of the matrix element $V_{cs}$ with dimuons 
have been reported by the CDHS, CCFR and CHARM2 Collaborations. 
Their results have been combined in a review paper 
of 2000 \cite{bargi} giving V$_{cs}$= 1.04$\pm$0.16. 
This value can be compared with the world 
average V$_{cs}$=1.023$\pm$0.036 computed by the 
Particle Data Group \cite{datagr1}.
\end{itemize}

\section{ Multimuon events and associated charm production}
\label{multi}

\par
\begin{figure}[hbtp!]
\begin{center}
\epsfig{figure=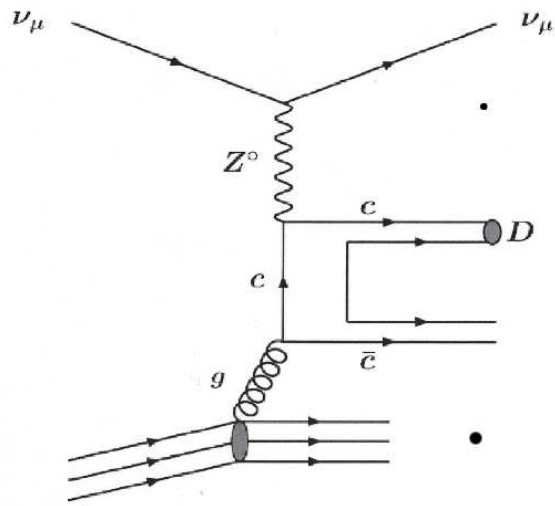,angle=-90,width=8.cm}
\caption {Gluon-boson fusion graph, NC reaction}
\label{fus}
\end{center}
\end{figure}

\begin{figure}[hbtp]
\begin{center}
\epsfig{figure=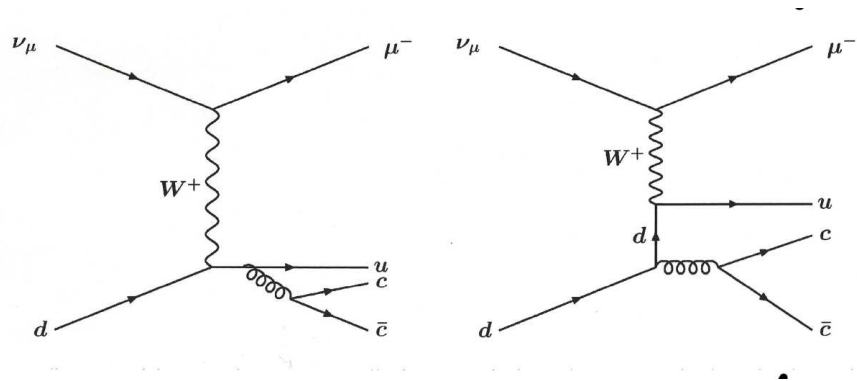,angle=-90.,width=12.cm}
\caption {Gluon-quark bremsstrahlung graph, CC reactions}
\label{brem}
\end{center}
\end{figure}

 The main channel for  open charm production in neutrino interaction is the CC 
reaction described in section 3 and experimentally studied with dimuons as 
described 
in section 5. However, many experiments have also searched for the rarer 
process of 
associated production of open charm. Theoretically $ c\overline{c}$ production is 
expected both in NC and CC processes. The main contribution should come from the 
so-called `` gluon-boson fusion `` process \cite{Gluck} shown in fig \ref{fus} in 
NC reactions, and from the `` gluon-quark bremsstrahlung `` process \cite{Hagiw} shown 
in fig \ref{brem} in CC reactions. 


Both processes are rare, mostly because of the higher threshold of 3.7 GeV for 
two charmed mesons, and difficult to isolate. Many searches for $c\overline{c}$ 
production were performed using the massive electronic detectors and looking for 
muons from charm decay. The first evidence only came in year 2000 from the high 
statistics, high energy NuTeV-E815 experiment at Fermilab. Its result was then 
confirmed by the CHORUS experiment at CERN, by detecting the decay of charmed 
hadrons in its emulsion target. In this section we recall the main results 
obtained from the multimuon searches with electronic detectors. The results 
from emulsion experiments are given in the next section. 

\subsection{Wrong-sign muons}

\par A signature for $c\overline{c}$ production in NC interactions is given 
by the so-called wrong-sign muon events. These are events in which the 
single muon in the final state has a lepton number opposite to that of 
the incident neutrino. The event is a NC interaction since the muon
with the incoming neutrino lepton number is missing. The wrong-sign muon is then 
attributed to the decay of a charmed hadron. Since there is no single 
charm production in NC interactions, it is assumed that a second undetected 
charmed hadrons is present in the hadron shower.
\par First searches for wrong-sign muons only led to negative results. 
For instance, data taken with the CCFR detector at Fermilab are analyzed in a 1989
paper \cite{mishra} resulting in a number of wrong sign muons compatible  with
that expected from background. First evidence for $c\overline{c}$ production, 
and up to now the only evidence with wrong-sign muon events, was obtained in 
year 2000 by the NuTeV-E815 experiment at Fermilab. The experiment running in 
a narrow-band beam with an average neutrino energy of 154~GeV quotes the 
following cross-section \cite {charmneutra}:
\par $\sigma( \numu N \rightarrow  c\overline{c}X)$=0.21$^{+0.18}_{-0.15}$ fb.
\par The existence of the process was proved, although the errors were 
too large to allow a quantitative analysis of the signal.

\subsection{Same-sign dimuons}

\par Same-sign dimuons can be a signature for $c\overline{c}$ production 
in CC reactions. For this to be true, it is required the presence of a 
leading muon with the sign corresponding to the lepton number of the 
incoming neutrino and of a second muon within the hadron shower, with 
the same sign of the leading muon. The second muon is associated to 
the decay of a charmed hadron. It is then assumed that a second, undetected,
charmed hadron is present, since in the case of single charm production
the muon from charm decay would have opposite sign with respect to the
leading muon. 
\par Several experiments have observed same-sign dimuon events, but no 
experiment could provide clearcut evidence for $c\overline{c}$ production. 
The statistics collected by the more important experiments are shown 
in table \ref{sames}.

\begin{table}[htbp]
\begin{center}
\begin{tabular}{|c|c|c|c|c|c}\hline

Experiment & $\mu_{++}$ & $\mu_{--}$ &ref&year\\\hline
CDHS & 52 & 74& \cite{degro}&1979\\
CHARM &52  & 74 &\cite{jonke}&1982\\
CCFR & 25 & 220 & \cite{sandler}&1993\\
NuTeV& 15&  101&\cite{schumm}&1988\\\hline
\end{tabular}
\caption {\large{Same-sign dimuons}} 
\label{sames}
\end{center}

\end{table}
\par Because of many different background processes contributing to those 
samples, no firm conclusion on the interpretation of these events has been 
reached. We quote as an example the statement made in the NuTeV paper \cite{schumm}:
{\sl The small excess of $\mu_{-}\mu_{-}$ is consistent with but does not 
require
$c\overline{c}$  production.}

\subsection {Trimuons}

\par Events with three muons in the final state can indicate $c\overline{c}$ 
production in CC reactions, similarly to the same-sign dimuons described above. 
The signature is cleaner, but the rates are further  reduced because both 
charmed hadrons are required to decay into a muon. Again, several experiments 
have collected a sample of few trimuons events, but none of them was 
able to prove $c\overline{c}$ production.  The statistics collected 
are given in table \ref{trimu}.

\begin{table}[htbp]
\begin{center}
\begin{tabular}{|c|c|c|c|c}\hline

Experiment & $\mu_{--+}$ or $\mu_{++-}$ &ref&year\\\hline
HPFW& 39& \cite{benve6}&1978\\
FNAL&    3  &\cite{barish}&1977\\
Gargamelle& 10  &\cite{garga2}&1982\\
CDHS & 2 &  \cite{trimucdhs}&1978\\
CHORUS& 42& \cite{trimu}&2003\\\hline
\end{tabular}\label{trimu}

\caption {\large{Results on trimuon detection}}
\label{trimu}
\end{center}
\end{table}

The largest statistics is that of the emulsion experiment CHORUS, which 
used for this analysis its hadron calorimeter as target for 
the neutrinos. In their paper \cite{trimu} the authors attribute 
the 42 events to various processes, without requiring a contribution 
from $c\overline{c}$ production.

\vskip 3cm

\section{Emulsion experiments}
\label{emul}
\par Four neutrino experiments have investigated charm physics using nuclear 
emulsions as target and detector. The spatial resolution of nuclear emulsions 
makes possible the direct detection of charmed particles through the observation 
of the few millimeters long decay-paths. Contrary to the electronic 
experiments described in the previous sections, the charmed hadrons detected 
in emulsions are then practically background free and so single events 
can give relevant information. In fact, following the proposal of 
Conversi \cite{conversiprop} two fundamental experiments, E247 at 
Fermilab and WA17 at CERN were carried on in the late seventies. The 
two experiments made use of an emulsion target followed by few electronic 
detectors to identify and measure momenta of charged particles. In 1976 
E247 \cite{Burhop,Read} was the first to measure the lifetimes of the 
new hadrons showing that these were consistent with the expectation for 
hadrons with charm. In 1979 the few events collected by WA17 
\cite{conversi1,conversi2} allowed to establish a difference in 
lifetime between charged and neutral charmed hadrons, confirming theoretical 
predictions for the decay process involving the c quark. WA17 also observed 
the decay of one charmed baryon \cite{Allas4}. 

\par After the two historical experiments, emulsion target were 
rarely used for the study of charm physics with neutrino beam, 
mainly because the analysis of emulsion film has to be performed manually at 
microscopes and the scanning phase makes it long and difficult to collect 
large statistics. Nevertheless, few important measurements were 
reported in the eighties by the E531 experiment at Fermilab. Ten years later 
a large sample of charm events was collected by the CHORUS 
experiment at CERN, which exploited the huge progress in the automation 
of the process of emulsion scanning. We shall now give some detail on 
these two experiments.

\subsection{E531}
\par The E531 experiment at Fermilab has collected in the early 
eighties a total of 3855 neutrino interactions and identified 121 
decays of charmed hadrons. As shown in Table \ref{e531dec}, E531 
was able to identify the different charmed hadrons and to provide 
evidence for the production, besides 
of D$^{0}$ and D$^{+}$, also of D$^{+}_{s}$ and $\Lambda_{c}^{+}$. 

\begin{table}[htbp] 
\begin{center}
\begin{tabular}{|c|c|}\hline particle &number\\ 
D$^{0}$ & 57\\ D$^{+}$ & 41\\ D$^{+}_{s}$&6\\ $\Lambda_{c}^{+}$&14\\\hline 
\end{tabular}
\caption {E531 charmed hadrons}
\label{e531dec}
\end{center} 
\end{table}

\par E531 described in  \cite{e531c} characteristics of the 
charm production by neutrinos and reported in  \cite{e531} 
the measurement of the ratio of 
cross-sections, $\sigma(charm)$ /$\sigma(total)= 5.4 \pm 0.7 \%$.
That value, although less precise than the measurements performed by 
electronic experiments using dimuon events, does not depend from 
B$_{\mu}$, 
the average branching ratio for charmed hadrons into muons. 

\subsection{CHORUS}
\par The CHORUS experiment, shortly described in section 4, took data 
in the wide-band neutrino beam at the CERN SPS from 1994 to 1997 and identified 
 the decay of 2059 charmed hadrons
in the 770 kg emulsion target. The collection of such a huge sample was made possible by the 
automation of the scanning with computer driven microscopes, and by the 
use of a new analysis technique, the so-called 'netscan method', 
originally developed for the DONUT experiment and described in 
 \cite{kodama}. The 'netscan method' uses all the 
micro-tracks which at the scanning level were found by joining emulsion 
grains, and reconstructs primary and secondary vertices and tracks 
of the neutrino interactions. 
\par The 2059 events with hadron decays were classified following 
the topology at the decay vertex, as Vn, for neutral charms, and as 
Cn for charged, where n is the number of charged tracks from the 
decay vertex. The number of events for each topology is 
given in Table \ref{charmtable}.

\begin{table}[htbp]
\begin{center}
\begin{tabular}{|c|c|}\hline topology &number\\
C1 & 461\\
V2 & 841\\
C3&501\\
V4&230\\
C5&23\\
V6& 3\\\hline
\end{tabular}
\end{center}
\caption {CHORUS  topologies of charm decays}
\label{charmtable}
\end{table}

\par The analysis has lad at various results. About 15 papers 
were published from the first 1998 \cite{annis} to the last in 2011 \cite{last}.
We list in the following the 
most relevant studies.

\begin{itemize}

\item
\par\underline{Cross-sections}
\par The E531 measurement of the total neutrino cross-section 
for charm production, independent from B$_{\mu}$, was repeated by CHORUS 
and completed with the measurement of the antineutrino cross-section \cite{anti}.

\item
\par \underline{ The production of the $\Lambda_{c}$}
\par CHORUS had limited capabilities of particle identification, nonetheless 
it succeeded to identify a sample of $\Lambda_{c}$ on statistical basis, 
by using the different flight lengths of  $\Lambda_{c}$ and D particles. 
The total neutrino cross-section for $\Lambda_{c}$ production relative to 
the CC cross-section was measured to be
\par $\sigma(\Lambda_{c})/\sigma(CC)$= $(1.54 \pm 0.35 \pm 0.18)$ x 
$10^{-2}$
\cite{lambda}
\par This result indicates that about 40$\%$ of all charmed hadrons produced 
in neutrino interaction at an average neutrino energy of 27 GeV are $\Lambda_{c}$. 
The production mechanisms of the $\Lambda_{c}$ were also investigated and, 
using topological and kinematical criteria, it was possible to separately
measure the contribution of the quasi-elastic process obtaining:

\par $\sigma (\Lambda_{c}(QE)) / \sigma(CC) = 0.23 ^{+.12}_ {-.06}$ x $10^{-2}$ \cite{quasi}.

\item
\par \underline{Associated charm production}
\par The search for $c\overline{c}$ associated charm production with 
electronic detectors has been discussed in sect \ref{multi}, where 
we recalled that the only positive signal was found in NC events by 
the NuTeV collaboration. CHORUS has directly searched for $c\overline{c}$ 
production by looking for events with two charmed particles decays. 
Double charm production in neutrino NC interactions was confirmed 
by the observation \cite{assoc2} of three double-decay events, 
corresponding to the following cross-sections ratio:

\par $\sigma$ ($c\overline{c}_{NC}$) / $\sigma(NC)$
=(3.62$^{+2.46}_{-2.42}$ (stat) $\pm$ 2.0 (syst) x 10$^{-3})$.
\par The experiment also found a candidate consistent with  double charm 
production in CC interaction, but preferred to quote just the upper limit
\par $\sigma$ ($c\overline{c}_{CC}$) / $\sigma(CC)$
 $\leq$ 9.69 x 10$^{-4}$.    

\item
\par \underline{D$^{0}$ decay branching ratios}
\par CHORUS has studied in detail the topological branching ratios 
of the {D$^{0}$} meson, an interesting subject given that 
at the time of publication (2005) only 64$\%$ of the {D$^{0}$} 
branching ratios had been measured. The results were given for the different
charm decays topologies in terms of 
number of prongs, i.e. the number of charged particles from the 
decay vertex. CHORUS measured the rates of the 2,4,6~prongs decays 
and used as normalization factor the PDG value \cite{datagr3} for the
4-prongs decay,  
$\Gamma(4prongs)/\Gamma(total)$ =.146 $\pm$0.005. 
With that normalization, by subtraction, also the 0-prong b.r. was obtained. 
The published results \cite{D02} are the following:

\par $\Gamma(0prongs)/\Gamma(total)$ = 0.218 $\pm$ 0.049 $\pm$ 0.036
\par $\Gamma(2prongs)/\Gamma(total)$ = 0.647$\pm$ 0.049 $\pm$ 0.031
\par $\Gamma(6prongs)/\Gamma(total)$ = (1.2 $^{+1.3}_{-0.9}$ $\pm$0.2) x 10$^{-3}$.

\item
\par \underline{Muonic branching ratios}
\par B$_{\mu}$, the inclusive branching ratio of charmed hadrons 
for decays with a muon, is an important normalization factor for 
the charm physics studied by the electronic detectors using 
dimuon events. CHORUS has performed an independent measurement of 
B$_{\mu}$ analyzing the muonic decay associated to the various decay 
topologies. The results \cite{topo} are summarized in table \ref{Bmu}, 
which gives for each topology background and efficiency, and the 
resulting corrected B$_{\mu}$.

\begin{table}[htbp]
\begin{center}
\begin{tabular}{|c|c|c|c|c|}\hline
 topology&selected&background ($\%$)&efficiency  &B$_{\mu}$\\
C1 &20&  0.8& 36.0 $\pm$3.4& 10.8 $\pm$2.4 $\pm$0.5\\
C3   &17&  8.4&26.4 $\pm$2.6 &6.1 $\pm$1.6 $\pm$0.6\\
V2+V4 &36&  9.8&30.1 $\pm$1.5 &8.1 $\pm$1.5 $\pm$0.3\\\hline
\end{tabular}
\caption{muon branching ratios and values of B$\mu$ }
\label{Bmu}
\end{center}
\end{table}

\par From the results given in table \ref{Bmu} an average 
$\overline{B}_{\mu}$ of (8.5 $\pm$0.9$\pm$0.6) x 10$^{-2}$
has been derived.

\end{itemize}
\vskip 3cm

\section{Conclusion}
\label{conclu}

Our recollection has summarized the many important results on
 properties of the c quark and of charmed hadrons collected
 in thirty years of experimentation with neutrino beams, together
 with a brief description of the techniques adopted. Nowadays
 experiments with neutrino are designed to study oscillations,
 aiming at a complete understanding of the pattern of neutrino masses
 and mixings, and perhaps of CP violations in the neutrino
 sector (a review on neutrino oscillation experiments can be
 found in \cite{Dore}). So, after having played an important
 role in the measurement of the CKM matrix, neutrino experiments are
 now engaged in the  new task of unveiling the  complete 
 Pontecorvo-Maki-Nakagawa-Sakata matrix. 
 \par We should add that in our opinion, charm physics with neutrinos
  has strongly influenced the 
 evolution of the overall picture of quarks in particle physics.
 By studying the role of valence and virtual quarks, the role of
 quark masses in the scattering and hadronization processes, neutrino
 charm physics has contributed to our current understanding of quarks
 in the Standard Model.

\vskip 1.cm
{\Large{Aknowledgments}}
\vskip 0.3cm  
\par
 We gratefully aknowledge the enlightening contribution of P.F.Loverre
in discussing and revising many different aspects of the paper,
and the critical reading of L.~Ludovici and B.~Saitta.

\bibliographystyle{unsrt}
\bibliography{charmlo}

\end{document}